\documentclass[cameraready]{Interspeech}

\title{Parameter-Efficient Continual Learning for Automatic Speech Recognition}

\author[affiliation={1}, orcid=0000-0003-3658-3925, correspondingauthor]{Steven}{Vander Eeckt}
\author[affiliation={1}, orcid=0000-0003-1331-5186]{Hugo}{Van hamme}

\address{
    $^1$ Department Electrical Engineering ESAT-PSI, KU Leuven, Leuven, Belgium 
}

\email{steven.vandereeckt@esat.kuleuven.be, hugo.vanhamme@esat.kuleuven.be}


\usepackage{comment}
\usepackage{amsmath,amsfonts}
\usepackage{algorithmic}
\usepackage{algorithm}
\usepackage{array}
\usepackage{textcomp}
\usepackage{stfloats}
\usepackage{url}
\usepackage{verbatim}
\usepackage{graphicx}
\usepackage{cite}
\usepackage{booktabs}
\usepackage{threeparttable}
\usepackage{multicol, multirow}
\usepackage{bm}
\usepackage{subfigure}

\usepackage{enumitem}

\def\D{{\cal D}}

\def\T{{\cal T}}

\begin{document}

\maketitle

\begin{abstract}
Speech foundation models enable strong general-purpose ASR and are attractive for downstream adaptation. However, their size and the catastrophic forgetting induced by sequential fine-tuning demand parameter-efficient and regularized training methods, motivating parameter-efficient continual learning (PECL). While PECL has been widely studied in NLP and vision, it has received less attention in ASR. 
In this paper, we propose a simple yet effective PECL method based on recent advances in parameter-efficient fine-tuning for ASR. We partition pretrained weight matrices into head and tail subspaces according to singular values and restrict adaptation to approximate rotations within the low-energy tail subspace, preserving dominant components and reducing forgetting. For subsequent tasks, rotations are combined via weight averaging to further improve retention. Experiments on two benchmarks demonstrate reduced forgetting and superior overall performance compared to recent PECL baselines.
\end{abstract}

\section{Introduction}
Automatic Speech Recognition (ASR) has undergone remarkable progress over the past decade. More recently, speech foundation models trained on massive and multilingual datasets have emerged as powerful general-purpose models \cite{whisper, owsm}. These models capture general speech knowledge, making them attractive for adaptation to specific downstream tasks.

Adapting such large-scale models, however, presents two fundamental challenges. First, their size--often above hundreds of millions of parameters--makes full fine-tuning computationally expensive. Parameter-efficient fine-tuning (PEFT) methods address this issue by updating only a small subset of parameters while keeping the rest of the model frozen. Second, na\"ively fine-tuning a model on new tasks leads to catastrophic forgetting (CF) \cite{catastrophicforgetting}, where performance on previously learned tasks deteriorates, which Continual learning (CL) aims to mitigate.

Speech foundation models must thus be adapted efficiently, and this adaptation must not lead to forgetting of previous tasks. This setting naturally gives rise to \emph{parameter-efficient continual learning} (PECL) \cite{zhao2024sapt}, where the goal is to sequentially adapt a large pretrained ASR model to (multiple) downstream tasks under strict parameter constraints, while preserving performance on earlier tasks without access to their training data.

In other domains, PECL has received considerable attention in recent years, particularly in Natural Language Processing (NLP) for adapting Large Language Models \cite{zhao2024sapt, corda, milora, oplora, osft, qiao2026merge} and in image classification \cite{chitale2023task, bilora, He_2025_CVPR, muralidhara2025clora, luo2026keeplora, ewclora}. Many of these approaches build upon low-rank adaptation (LoRA) \cite{lora}, for example by constraining its update subspace \cite{bilora, oplora, luo2026keeplora}, averaging task-specific adapters \cite{chitale2023task, qiao2026merge}, or modifying its initialization \cite{corda, milora}.

In ASR, however, PECL has been studied only to a limited extent. \cite{xu24h_interspeech} adapts Whisper \cite{whisper} using orthogonal LoRA \cite{wang-etal-2023-orthogonal} and AdaLoRA \cite{adalora} to reduce forgetting and allocate parameters across layers, while \cite{ugan25_interspeech} applies weight averaging \cite{weight_averaging} to LoRA modules. Both approaches, however, do not explicitly address forgetting with respect to the initial pretrained model. Task-specific adapter \cite{adapters} approaches have also been explored \cite{eeckt_adapters}, but require task identity at inference time. More broadly, continual learning in ASR has primarily focused on storing past data \cite{lifelongasr, eeckt2021continual, svr} or regularizing the training \cite{updating_only, weight_averaging, wang23d_interspeech, vanderEeckt2026inversehessian}, without explicitly considering parameter-efficiency constraints.

Recently, \cite{wang2025ssvd} introduced structured Singular Value Decomposition (SSVD) as a PEFT method for ASR. SSVD decomposes weight matrices via SVD and adapts the top $k$ singular directions through a learned rescaling and approximate rotation, showing strong performance for PEFT in ASR.

Building on SSVD, we propose a simple yet effective PECL method for ASR. In contrast to SSVD, we retain only the rotation component and partition each pretrained weight matrix into a \textit{head} and \textit{tail} according to decreasing singular values. Adaptation is restricted to the low-energy \textit{tail} subspace, leaving the dominant singular directions untouched and thereby reducing interference with previously learned tasks. For subsequent tasks, we combine rotations via weight averaging to further mitigate forgetting. We evaluate the proposed method on two benchmarks and compare against a broad range of PECL approaches. Our results demonstrate consistently reduced forgetting and improved overall performance.

Our contributions are threefold: 
(i) we provide the most extensive empirical study of PECL for ASR to date, implementing and evaluating recent PECL methods originally proposed for NLP and vision; 
(ii) we introduce a novel PECL method tailored to ASR that achieves stronger retention and superior performance compared to these baselines; 
(iii) we present an ablation study that identifies the key components underlying the effectiveness of our method.

\section{Problem Formulation}

Let an initial model with parameters $\bm{\theta}^0 \in \mathbb{R}^N$ be trained on an initial set of tasks $\mathcal{T}_0$, whose data $(\bm X, \bm y) \in \D_0$, with $\bm X \in \mathbb{R}^{F\times d_s}$ the input utterance (consisting of $F$ frames of dimension $d_s$) and $\bm y \in \mathbb{R}^w$ the corresponding set of $w$ ground truth tokens, is no longer available. In parameter-efficient continual learning (PECL), this model is sequentially adapted to a stream of new tasks $\mathcal{T}_1, \dots, \mathcal{T}_t$, with each task $\T_i$ consisting of labeled data $(\bm X, \bm y) \in \D_i$. To learn task $\mathcal{T}_i$ from $\mathcal{D}_i$, the parameters $\bm{\theta}^{i-1}$ are updated to $\bm{\theta}^i$ in a parameter-efficient manner, while retaining performance on previous tasks $\{\T_j\}_{0 \leq j \leq i-1}$.

The objective of PECL is twofold: (i) effective adaptation to new task $\T_i$ under strict parameter-efficiency constraints, and (ii) preservation of performance on all previously learned tasks $\mathcal{T}_0, \dots, \mathcal{T}_{i-1}$ without access to their (full) data. We restrict adaptation to the weight matrices of linear layers, keeping all other parameters, including the output layers, frozen. Task identity is assumed unavailable at inference time.

\section{Our Method: Continual SSVD}
\begin{figure}
    \centering
    \includegraphics[width=\linewidth]{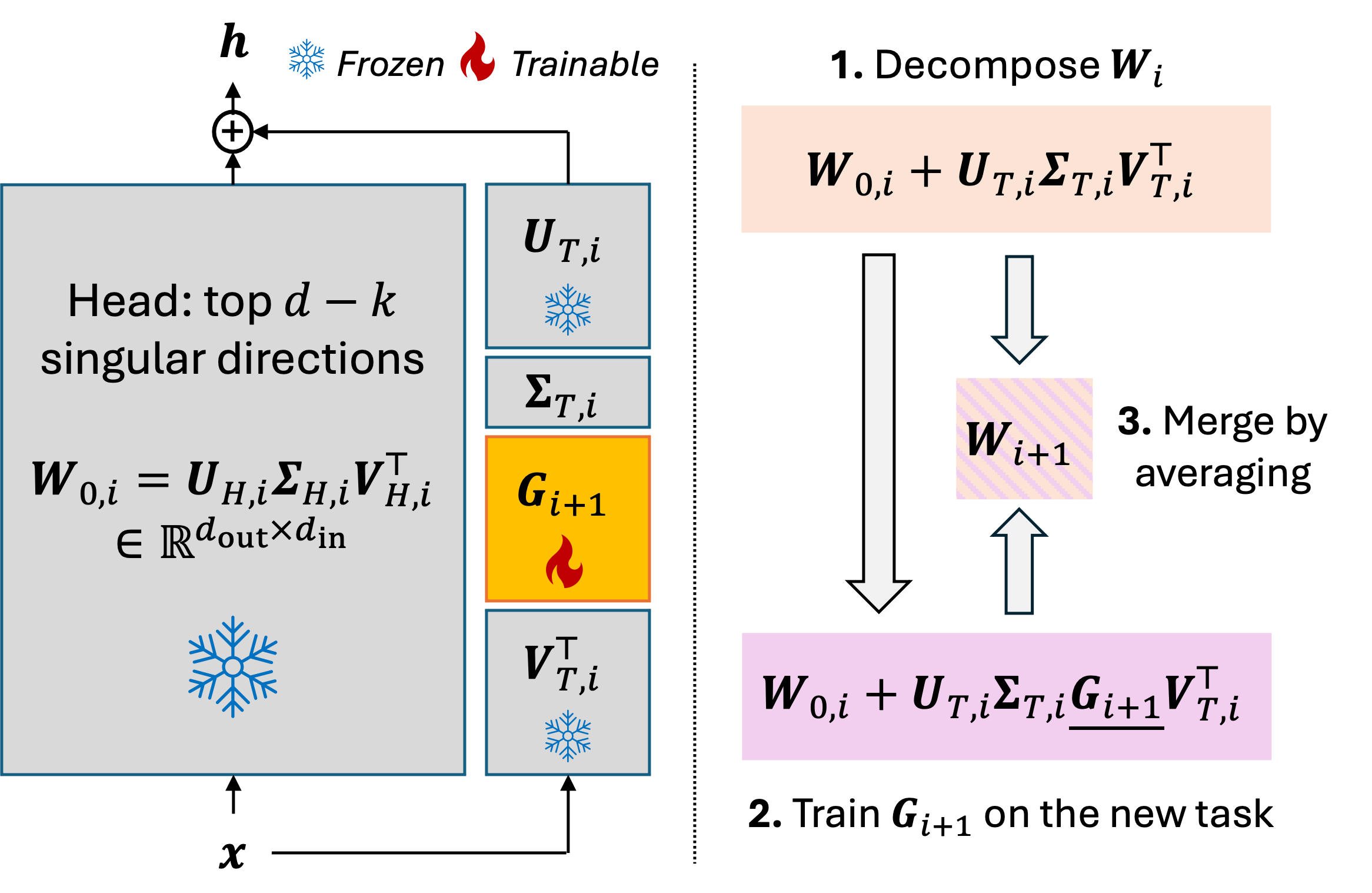}
    \caption{Overview of CSSVD for a single linear layer (bias omitted) when learning task $\mathcal{T}_{i+1}$. 
Right: (1) current weights $\bm W_i$ are decomposed via SVD into \textit{head} and \textit{tail} subspaces; 
(2) an approximate rotation matrix $\bm G_{i+1}$ is introduced within the \textit{tail} and optimized on the new task to obtain $\tilde{\bm W}_{i+1}$; 
(3) $\bm W_i$ and $\tilde{\bm W}_{i+1}$ are merged through averaging to produce $\bm W_{i+1}$. 
Left: schematic view of the decomposition for an input $\bm x \in \mathbb{R}^{d_\text{in}}$ and output $\bm h \in \mathbb{R}^{d_\text{out}}$, with only $\bm G_{i+1}$ in the \textit{tail} trainable.}
    \label{fig:overview}
\end{figure}
\subsection{Learning the first task}
\label{sec:first_task}
Let $\bm W \in \mathbb{R}^{d_\text{out} \times d_\text{in}}$ be the weight matrix of a linear layer of the initial model $\bm \theta^0$ whose output is $\bm h =\bm W\bm x$ for an input $\bm x \in \mathbb{R}^{d_\text{in}}$ (we omit the bias $\bm b \in \mathbb{R}^{d_\text{out}}$ for simplicity). Its singular value decomposition (SVD) is given as:
\begin{equation}
    \bm W = \bm U \bm \Sigma \bm V^\top
\end{equation}
with $\bm U \in \mathbb{R}^{d_\text{out} \times d}$ and $\bm V \in \mathbb{R}^{d_\text{in} \times d}$, the left and right singular vectors, resp.,  and $\bm \Sigma \in \mathbb{R}^{d \times d}$ a diagonal matrix with the singular values sorted in decreasing order, with $d=\min(d_\text{in}, d_\text{out})$. 

To adapt $\bm W$ in a parameter-efficient manner to new tasks, SSVD (structured SVD) \cite{wang2025ssvd} proposes to introduce and learn a rotation matrix $\bm G \in \mathbb{R}^{k \times k}$ and rescaling $\Delta \bm \Sigma \in \mathbb{R}^k$ for the top $k$ singular directions (associated with highest singular values) from $\bm U \bm \Sigma \bm V^\top$, with $k=pd$ with $p \in (0, 1)$ a hyper-parameter. In the \textit{approximate orthogonal constraint} version, they define $\bm G:=\bm I-2\bm K$ with $\bm I$ the $k \times k$ identity matrix and $\bm K \in \mathbb{R}^{k \times k}$ a skew symmetric but otherwise unconstrained matrix. The rescaling and rotation thus require learning $k$ and $k(k-1)/2$ parameters, resp., totaling $k+k(k-1)/2$ parameters. 

While SSVD adapts the $k$ directions associated with largest singular values, we operate on the $k$ smallest singular directions to reduce interference between new task $\mathcal{T}_1$ and initial tasks $\mathcal{T}_0$.

Referring to the top $d-k$ directions (associated with largest singular values) as the \textit{head}, and to the bottom $k$ directions as the \textit{tail}, we can write the decomposition of $\bm W$ as follows:
\begin{equation}
    \bm W =  \bm U_H \bm \Sigma_H \bm V_H^\top + \bm U_T \bm \Sigma_T \bm V^\top_T = \bm W_0 + \bm U_T \bm \Sigma_T \bm U_T^\top
\end{equation}
To obtain $\bm W_1$, the weight matrix adapted to task $\mathcal{T}_1$, we apply SSVD within the \textit{tail} by introducing an approximate rotation matrix $\bm G_1$. Since we adopt the approximate orthogonality constraint of SSVD, $\bm G_1$ is not strictly constrained to be orthogonal and can implicitly capture both rotation and rescaling. We therefore omit the explicit rescaling term $\Delta \bm \Sigma_1$ and introduce only $\bm G_1$. The adapted weight matrix is then given by:
\begin{equation}
    \bm W_1 = \bm W_0 + \bm U_T \bm \Sigma_T \underline{\bm G_1} \bm V_T^\top
\end{equation}
where $\bm G_1 = \bm I-2\bm K_1$ with initially $\bm K_1 \approx 0$, and only the underlined part is trained. Since the update is restricted to the tail subspace while the head remains fixed, interference with initial tasks $\T_0$ is reduced.

\subsection{Learning additional tasks}
\label{subsec:task2}
To learn a new task $\T_2$, we start from the same decomposition. Assuming the training of $\bm G_1$ during the first task has not altered the division between \textit{head} and \textit{tail}, the decomposition can be written as follows:
\begin{equation}
    \tilde{\bm W}_2 = \bm W_{0} + \bm U_T{\bm \Sigma}_T\underline{\bm G_2} \tilde{\bm V}_T^\top
\end{equation}
Since $\tilde{\bm V}_T = \bm V_T \bm G_{1}^\top$, the formulation for $\tilde{\bm W}_2$ becomes:
\begin{equation}
        \tilde{\bm W}_2 = \bm W_0 + \bm U_T {\bm \Sigma}_T\underline{\bm G_2} \bm G_1{\bm V}_T^\top
\end{equation}
Introducing $\bm G_2$ may induce forgetting of task $\T_1$. To mitigate this, we merge $\bm G_1$ and $\bm G_2 \bm G_1$ via averaging, a strategy previously shown effective for continual learning in ASR \cite{weight_averaging}. The resulting weight matrix $\bm W_2$ is given by:\begin{equation}
\bm W_2 = \bm W_0 + \bm U_T\bm\Sigma_T \left((1-\alpha)\bm G_1 
+\alpha\bm G_2\bm G_1 \right)\bm V_T^\top
\label{eq:rot_avg}
\end{equation}
with $0 \leq \alpha \leq 1$. Note that this solution $\bm W_2$ corresponds to computing a convex combination of the task $\T_1$ solution $\bm W_1$ and the
task $\T_2$ solution $\tilde{\bm W}_2$:
\begin{equation}
    \bm W_2 = (1-\alpha)\bm W_1 + \alpha \tilde{\bm W}_2
    \label{eq:w_avg}
\end{equation}

\subsection{Practical Implementation}
\label{sec:adaptive_split}

The formulation in Eq.~\eqref{eq:rot_avg} assumes that the partition into 
\textit{head} and \textit{tail} remains fixed across tasks. 
In practice, however, after completing task $\T_{i}$ we only store the 
resulting weight matrix $\bm W_{i}$. 
Before learning a new task $\T_{i+1}$, we recompute its SVD and define a new head and tail by selecting again the $d-k$ largest and 
$k$ smallest singular directions, respectively. 
The adaptation for $\T_{i+1}$ is then restricted to the new tail:
\begin{equation}
    \tilde{\bm W}_{i+1} = \underbrace{\bm U_{H,i} \bm \Sigma_{H, i} \bm V_{H, i}^\top}_{\bm W_{0,i}}
    + \bm U_{T,i} \bm \Sigma_{T,i} \underline{\bm G_{i+1}} \bm V_{T, i}^\top
\end{equation}

This re-computation of \textit{head} and \textit{tail} allows singular directions whose importance has 
increased during tasks $\{\T_j\}_{j<i}$ to move from the tail to the head.  Highly important directions thus become protected from further modification, while adaptation capacity remains concentrated in the low-energy subspace.

When no singular value crossings occur between head and tail, the above formulation reduces to the \textit{fixed} case from Sec. \ref{subsec:task2} up to a change of basis within 
the tail subspace. However, as the partition may change, Eq.~\eqref{eq:rot_avg} does not generally apply. 
Therefore, in practice we compute the merged solution directly via  Eq.~\eqref{eq:w_avg}, using $\alpha=1/(i+1)$ following \cite{weight_averaging}.

Figure \ref{fig:overview} summarizes Continual SSVD (CSSVD).

\section{Experiments}

\begin{table*}
\centering
\begin{threeparttable}
\caption{Results of the experiments. Tasks are learned from left to right; WERs are measured after learning all tasks. \textit{Params} indicates number of trainable parameters. Best result per column by PECL methods is in \textbf{bold}. Negative BWT indicates forgetting.}
\setlength{\tabcolsep}{4pt}
\begin{tabular}{
l r c@{\hspace{5pt}} c@{\hspace{5pt}} c@{\hspace{5pt}} c@{\hspace{5pt}} c@{\hspace{5pt}} c@{\hspace{5pt}} c@{\hspace{5pt}}
c@{\hspace{5pt}} c@{\hspace{5pt}} c@{\hspace{5pt}} c@{\hspace{5pt}} c@{\hspace{5pt}} c@{\hspace{5pt}} c@{\hspace{5pt}}
}
\toprule
& & \multicolumn{7}{c}{\textbf{\textit{Experiment 1}}} & \multicolumn{7}{c}{\textbf{\textit{Experiment 2}}} \\
\cmidrule(lr){3-9} \cmidrule(lr){10-16}
& & \multicolumn{5}{c}{\textbf{WER$\downarrow$ per task}} & \multicolumn{2}{c}{\textbf{Average}} & \multicolumn{5}{c}{\textbf{WER$\downarrow$ per task}} & \multicolumn{2}{c}{\textbf{Average}} \\
\cmidrule(lr){3-7} \cmidrule(lr){8-9} \cmidrule(lr){10-14} \cmidrule(lr){15-16}
\textbf{Method} & \textbf{Params} &
\textbf{ENG} & \textbf{DEU} & \textbf{ESP} &
 \textbf{NL} & \textbf{VL} &
\textbf{WER$\downarrow$} & \textbf{BWT$\uparrow$} & \textbf{ENG} & \textbf{DEU} & \textbf{ESP} &
 \textbf{VL} & \textbf{DVL} &
\textbf{WER$\downarrow$} & \textbf{BWT$\uparrow$}   \\
\midrule
Initial model & -- &
13.4 & 11.3 & 11.3 & 45.7 & 37.9 & 22.48  & -- & 13.4 & \phantom{1}11.3 & 11.3 & 37.9  & 86.0 & 31.98 & -- \\
Full Fine-Tuning & 244.8M & 24.3 & 57.3 & 21.7 & 27.7 & 13.6 & 28.94 & -18.2 & 45.9 & \phantom{1}91.5 & 43.4 & 34.2 & 29.7 & 48.95 & -41.0 \\
Separate Model & 244.8M  &
13.4 & 11.3 & 11.3 & 22.4 & 13.6 & 14.38 & \phantom{-3}0.0 & 13.4 & \phantom{1}11.3 & 11.3 & 15.2 & 29.7 & 16.17 & \phantom{-4}0.0 \\
\midrule
LoRA & 9.3M & 41.8 & 88.4 & 44.5 & 28.9 & 14.7 & 43.66 & -35.7 & 90.5 & 100.0 & 98.0 & 53.5 & \textbf{29.9} & 74.70 & -72.5 \\
LoRA + FTA & 9.3M & 16.4 & 20.6 & 15.1 & {27.2} & 19.0 & 19.64 & \phantom{1}-3.6 &  17.3 & \phantom{1}22.0 & 16.3 & \textbf{20.3} & 57.1 & 26.58 & \phantom{4}-4.8 \\
SSVD & 9.0M & 40.8 & 85.3 & 43.1 & 37.4 & 15.9 & 44.49 & -36.4 & 97.6 & 100.0 & 93.7 & 49.9 & 32.7 & 74.76 & -71.7 \\
MiLoRA & 9.3M & 47.4 & 91.7 & 48.5 & 30.2 & \textbf{14.1} & 46.39 & -39.6 & --- & --- & --- & --- & --- & --- & --- \\
OPLoRA & 9.3M & 33.8 & 81.6 & 36.3 & 30.2 & 15.7 & 39.53 & -30.3 & --- & --- & --- & --- & --- & --- & --- \\
BiLoRA & 9.3M & 19.4 & 34.7 & 17.8 & 30.0 & 16.0 & 23.58  & \phantom{1}-9.7 & 33.5 & \phantom{1}71.1 & 34.6 & 36.2 & 35.9  & 42.25   &  -30.4 \\
EWC-LoRA & 9.3M & 23.9 & 60.5 & 23.9 & \textbf{25.8}  & 17.0 & 30.21 & -18.6 &  30.7 & \phantom{1}72.9 & 30.0 & 23.2 & 43.7 & 40.06 & -26.1 \\
\midrule
CSSVD & 8.9M & \textbf{14.9} & \textbf{14.4} & \textbf{13.4} & {28.6} & 20.4 & \textbf{18.33}\tnote{a} & \phantom{1}\textbf{-1.9} & \textbf{15.3} & \phantom{1}\textbf{14.9} & \textbf{13.2} & 21.5 & 59.2 & \textbf{24.82}\tnote{a} & \phantom{4}\textbf{-2.2} \\
\bottomrule
\end{tabular}
\begin{tablenotes}
\footnotesize
\item[a] Significantly outperforms all PECL baselines and Full Fine-Tuning.
\end{tablenotes}
\label{tab:exp1}
\end{threeparttable}
\end{table*}

Experiments are done in ESPnet2 \cite{watanabe2018espnet}. More detailed information and code are available at our Github repository \footnote{https://github.com/StevenVdEeckt/pecl-for-asr}.

\noindent \textbf{Model.} We use Open Whisper-style Speech Model (OWSM) v3.2 small \cite{owsm}, comprising nine E-Branchformer \cite{e_branchformer} encoder and nine Transformer decoder layers.  A CTC branch is used only during training.  The model has a vocabulary of $50{,}000$ output tokens and contains 366.7M parameters in total.  It was pretrained on 180k hours of multilingual speech from 151 languages. During adaptation, we update only the weight matrices of the linear layers (excluding output layers). For all these linear layers, $d=\min(d_\text{out}, d_\text{in})=768$. Each method runs for 20 epochs using Adam \cite{adam}, with the learning rate selected based on the validation set of the new task on the first adaptation.

\noindent\textbf{Data.} We conduct two experiments. In both, English (ENG), German (DEU), and Spanish (ESP) from Common Voice \cite{commonvoice} serve as initial tasks $\T_0$, as they are included in OWSM v3.2's pretraining corpus \cite{owsm}. In Experiment~1, following \cite{svr}, we use Corpus Gesproken Nederlands (CGN) \cite{cgn}, split into Dutch from the Netherlands (NL) and Belgium (VL), yielding tasks $\T_1$ and $\T_2$, respectively. CGN covers diverse speech styles, including interviews, lectures, and broadcast recordings. In Experiment~2, VL serves as $\T_1$, and $\T_2$ comprises strongly dialectal Flemish speech from the corpus of Southern Dutch Dialects (GCND) \cite{gcnd}, referred to as DVL.

\noindent\textbf{Baselines}. 
We compare against the following methods and describe how each adapts linear layer $\bm W \in \mathbb{R}^{d_\text{out} \times d_\text{in}}$ to task $\T_{i+1}$:
\begin{enumerate}[label=(\alph*)]
    \item LoRA \cite{lora}: LoRA learns a low-rank adaptation of the form 
    $\bm W_{i+1} = \bm W_i + \bm B_{i+1}\bm A_{i+1}$, 
    where $\bm B_{i+1} \in \mathbb{R}^{d_\text{out} \times r}$ and 
    $\bm A_{i+1} \in \mathbb{R}^{r \times d_\text{in}}$, with $r$ a rank hyper-parameter. LoRA does not incorporate any explicit mechanism to mitigate CF.
    \item LoRA + FTA: We combine LoRA with Fine-Tuning with Averaging (FTA) \cite{weight_averaging}. After learning the low-rank update $\bm B_{i+1}\bm A_{i+1}$, the weights are updated as 
    $\bm W_{i+1} = \bm W_i + \eta \bm B_{i+1}\bm A_{i+1}$, 
    with $\eta = 1/(i+2)$. This setting is related to \cite{ugan25_interspeech}; however, we additionally include the pretrained model in the averaging, as excluding it would cause forgetting of $\T_0$.    
    \item SSVD \cite{wang2025ssvd}: The structured SVD method from Sec.~\ref{sec:first_task}.    
    \item MiLoRA \cite{milora}: A LoRA variant in which $\bm B_{i+1}$ and $\bm A_{i+1}$ are initialized using the $r$ singular directions associated with the $r$ smallest singular values of $\bm W_i$.
    \item OPLoRA \cite{oplora}: Based on the top $k_\text{OP}$ singular directions of $\bm W_i$ (associated with largest singular values), OPLoRA defines projection matrices $\bm P_L \in \mathbb{R}^{d_\text{out} \times d_\text{out}}$ and $\bm P_R \in \mathbb{R}^{d_\text{in} \times d_\text{in}}$ so that $\bm W_{i+1} = \bm W_i + \bm P_L \bm B_{i+1}\bm A_{i+1}\bm P_R$.  The update is constrained to lie in the orthogonal complement of the top $k_\text{OP}$ singular subspace of $\bm W_i$. Following \cite{oplora}, we set $k_\text{OP}=126$.
    \item BiLoRA \cite{bilora}: BiLoRA employs a {bilinear} update in fixed orthogonal bases, i.e., $\bm W_{i+1} = \bm W_i + \bm F_{\text{out}}\,\bm B_{i+1}\,\bm F_{\text{in}}^{H}$, where $\bm F_{\text{out}}$ and $\bm F_{\text{in}}$ are fixed 1D discrete Fourier transform matrices. While the original formulation assumes $\bm W \in \mathbb{R}^{d \times d}$, we use $\bm F_{\text{out}} \in \mathbb{C}^{d_\text{out}\times d_\text{out}}$ and  $\bm F_{\text{in}} \in \mathbb{C}^{d_\text{in}\times d_\text{in}}$, so that $\bm B_{i+1} \in \mathbb{C}^{d_\text{out}\times d_\text{in}}$. Task separation is achieved by enforcing sparsity in $\bm B_{i+1}$ (containing $k_\text{B}$ nonzero elements), with different tasks activating (approximately) non-overlapping coefficients.
    \item EWC-LoRA \cite{ewclora}: EWC-LoRA combines LoRA with Elastic Weight Consolidation (EWC) \cite{ewc}. Importance weights are computed from the diagonal of the Fisher information matrix on $\T_0$'s data at $\bm W_i$, and used to regularize the low-rank update $\bm B_{i+1}\bm A_{i+1}$. This requires access to pretraining data; we assume that only ENG data is available for computing importance weights, which are later accumulated across tasks. We use the regularization weight $\lambda_\text{EWC}$ reported in \cite{eeckt2021continual}.
\end{enumerate}
MiLoRA, OPLoRA, BiLoRA, and EWC-LoRA originate from domains outside ASR and, to our knowledge, have not yet been evaluated on ASR. Each PECL method trains approx. 9.0M parameters, corresponding to $r=20$ for LoRA-based methods and $p=0.40$ for (C)SSVD. We additionally report Full Fine-Tuning (FFT, applied to the same layers as PECL methods) and Separate Model (which uses task-specific model $\bm{\theta}^i$ from FFT to decode $\T_i$, assuming access to a task oracle) as references.

\noindent \textbf{Metrics.} 
We report word error rate (WER, in $\%$) for each task using the final model. \emph{Average WER}, our main metric, denotes the mean WER over all learned tasks. Backward Transfer (BWT) measures forgetting by comparing the final WER of each task to its WER immediately after training; it is defined as the average WER decrease on previous tasks, where negative values indicate forgetting. Statistical significance in Average WER is assessed using the Wilcoxon signed-rank test on per-utterance error counts \cite{Strik2000ComparingTR} at the 0.1\% level.

\section{Results}
Table \ref{tab:exp1} shows the results of both experiments.

\subsection{Experiment 1}

CSSVD achieves the lowest Average WER among all methods. 
While LoRA, SSVD, OPLoRA, and MiLoRA successfully learn the new tasks, they suffer from catastrophic forgetting, with the WER on previous tasks increasing by more than 30 points—exceeding FFT. 
For LoRA and SSVD, this behavior is expected, as neither method incorporates a mechanism to mitigate CF. 
MiLoRA shows that initialization alone is insufficient to preserve prior knowledge. 
OPLoRA reduces LoRA's forgetting by 15\%, but forgets more from NL ($\T_1$) when learning VL ($\T_2$), suggesting competition outside the protected top-$k_\text{OP}$ subspace. Increasing $k_\text{OP}$ further mitigates forgetting (Sec.~\ref{subsec:ablation}).

EWC-LoRA reduces LoRA's forgetting by $48\%$. Although initial importance weights are computed only on ENG, forgetting on ESP is reduced to a similar extent, indicating that EWC generalizes beyond the task used to estimate the Fisher information. Moreover, importance weights accumulated on NL ($\T_1$) help balance NL and VL when learning VL ($\T_2$). Stronger regularization may further mitigate forgetting, though at the cost of slower adaptation to new tasks, consistent with prior reports of mixed results for EWC-based CL in ASR \cite{lifelongasr, eeckt2021continual, ahadzi25_interspeech}.

Among the baselines, LoRA + FTA and BiLoRA perform best, reducing LoRA’s forgetting by 90\% and 73\%, respectively. 
BiLoRA achieves better performance on new tasks, whereas LoRA + FTA retains prior knowledge substantially better.

CSSVD, however, further improves upon both methods, reducing their Average WER by 7-22$\%$. Compared to SSVD, CSSVD reduces forgetting by 95\% while maintaining effective adaptation to new tasks. 
Relative to LoRA + FTA, it reduces forgetting by an additional 45\%, while incurring only a marginal increase in WER on the new tasks. 

\subsection{Experiment 2}
Exp.~2 confirms these findings but is more challenging, as DVL proves particularly difficult for OWSM. Adapting to DVL ($\T_2$) induces substantially more forgetting than in Exp.~1: LoRA and SSVD almost entirely forget ENG, DEU, and ESP ($\T_0$). CSSVD again achieves the best performance, reducing LoRA’s forgetting by 97\% and LoRA + FTA's by more than 50\%, the latter remaining the strongest baseline. However, in reducing forgetting, CSSVD—as well as LoRA + FTA—does not reach the low DVL ($\T_2$) WER of LoRA or SSVD. Consequently, approximating the Separate Model, which retrains all parameters and avoids forgetting, is harder than in Exp.~1. Nevertheless, CSSVD comes closest, improving the best baseline by 7\%. BiLoRA exhibits substantially more forgetting than in Exp.~1, while EWC-LoRA again balances VL ($\T_1$) and DVL ($\T_2$) reasonably well but suffers from catastrophic forgetting of $\T_0$, with similar degradation for ENG and ESP despite only ENG being used to estimate the initial importance weights.

\subsection{Ablation Study}
\label{subsec:ablation}
\begin{table}
    \centering
    \begin{threeparttable}
    \caption{Ablation study (on Exp. 1) of the proposed method. Rows marked with "$\rightarrow$" denote alternative variants.}
    \begin{tabular}{l l c@{\hspace{5pt}} c@{\hspace{5pt}}}
    \toprule
    & \multirow{2}{*}{\textbf{Model}} & \multicolumn{2}{c}{\textbf{Average}} \\
    \cmidrule(lr){3-4}
     & & \textbf{WER}$\downarrow$ & \textbf{BWT}$\uparrow$  \\
    \midrule
    1. & CSSVD & 18.33 & \phantom{2}-1.9 \\
    2. & $\rightarrow$ Keep initial head-tail separation & 18.40\tnote{b} & \phantom{2}-1.9 \\
    3. & $\rightarrow$ Do not average, i.e. skip Eq. \eqref{eq:w_avg} & 19.16\tnote{a} & \phantom{2}-3.3 \\
    4. & $\rightarrow$ Train rotation + rescaling & 18.27\tnote{b} & \phantom{2}-1.8 \\
    5. & SSVD + FTA & 19.22\tnote{a} & \phantom{2}-2.7 \\
    6. & OPLoRA [$k_\text{OP}=461$] & 31.61\tnote{a} & -20.2 \\
    \bottomrule    
    \end{tabular}
    \begin{tablenotes}
    \footnotesize
    \item[a] Significant deterioration with respect to the reference method.
    \item[b] No significant difference with respect to the reference method.    \end{tablenotes}
    \label{tab:ablation}
    \end{threeparttable}
\end{table}

Table~\ref{tab:ablation} presents an ablation, providing following observations:
\begin{itemize}
    \item Row~2 shows that keeping the initial head--tail separation and using Eq.~\eqref{eq:rot_avg} instead of Eq.~\eqref{eq:w_avg} yields nearly identical performance, with the latter offering a simpler implementation. 
    \item Row~3 demonstrates that skipping the averaging step substantially increases forgetting and Average WER. Nevertheless, this variant would still outperform all baselines in Table~\ref{tab:exp1}, indicating that restricting adaptation to the bottom $k$ singular directions is already highly beneficial. 
    \item Comparing Row~3 to SSVD in Table~\ref{tab:exp1} further highlights that adapting the lowest-$k$ singular directions, rather than the top-$k$ as in SSVD, forms the most critical component of CSSVD.
    \item Row~4 evaluates the effect of reintroducing explicit rescaling. The impact is negligible, confirming that the approximate rotation alone   suffices.
    \item Row~5 shows that combining SSVD with FTA significantly reduces forgetting compared to SSVD, but it still lags behind CSSVD and exhibits roughly 50\% higher forgetting, illustrating that averaging alone does not suffice.
    \item Row~6 increases $k_\text{OP}$ in OPLoRA so that its update spans a subspace comparable to CSSVD.  While this improves OPLoRA's performance, it remains inferior to CSSVD. This suggests that CSSVD's gains stem not only from restricting adaptation to the tail subspace, but also from the specific form of transformation permitted within that subspace.

\end{itemize}

\section{Conclusion}
We study parameter-efficient continual learning for ASR, a setting that has received considerable attention in NLP and vision but remains underexplored in ASR. Building on SSVD, a recent PEFT method for ASR, we propose CSSVD, a PECL approach that decomposes linear weight matrices into head and tail subspaces based on singular values and learns an approximate rotation within the tail. Leaving the dominant singular directions unchanged, CSSVD reduces interference with previously learned tasks. For subsequent tasks, we further mitigate interference within the shared tail subspace through weight averaging. Across two benchmarks, CSSVD achieves the best performance and consistently reduces forgetting compared to recent PECL methods from NLP and vision adapted to ASR. An ablation study confirms that its effectiveness stems from the combination of (i) restricting adaptation to the tail subspace, (ii) using approximate rotations as the adaptation mechanism, and (iii) applying averaging across tasks.

Currently, our method treats all layers uniformly. A promising direction for future work is to allocate adaptation capacity more selectively, prioritizing layers with high relevance for the new task and low interference with previous tasks.

\section{Acknowledgments}
Research supported by Research Foundation Flanders (FWO) under grant S004923N of the SBO programme.

\section{Generative AI Use Disclosure}
Generative AI tools were used to assist with minor language editing and phrasing improvements. All scientific content, experiments, and conclusions were developed by the authors, who take full responsibility for the manuscript.

\bibliographystyle{IEEEtran}
\bibliography{main}

\end{document}